\documentclass[fleqn,usenatbib]{mnras}

\usepackage{newtxtext,newtxmath}
\usepackage{CJK}
\usepackage{svg}

\usepackage[T1]{fontenc}

\DeclareRobustCommand{\VAN}[3]{#2}
\let\VANthebibliography\thebibliography
\def\thebibliography{\DeclareRobustCommand{\VAN}[3]{##3}\VANthebibliography}

\usepackage{graphicx}	
\usepackage{amsmath}	
\usepackage[final]{changes}
\usepackage{ulem}

\title[VSI with Partially Reflecting Boundaries]{
Vertical Shear Instability with Partially Reflecting Boundary Conditions}
\author[Wu, Yu \& Cui]{
Yuzi Wu (吴玉孜),$^{1,2}$
Cong Yu (余聪), $^{1,2}$\thanks{E-mail: yucong@mail.sysu.edu.cn}
Can Cui (崔灿) $^{3}$
\\
$^{1}$School of Physics and Astronomy, Sun Yat-Sen University, Zhuhai 519082, China\\
$^{2}$CSST Science Center for the Guangdong-Hong Kong-Macau Greater Bay Area, Zhuhai 519082, China\\
$^{3}$Department of Astronomy and Astrophysics, University of Toronto, Toronto, ON M5S 3H4, Canada \\
}

\date{Accepted XXX. Received YYY; in original form ZZZ}

\pubyear{2024}

\begin{document}
\begin{CJK*}{UTF8}{gbsn}
\label{firstpage}
\pagerange{\pageref{firstpage}--\pageref{lastpage}}
\maketitle

\begin{abstract}
The vertical shear instability (VSI) is widely believed to be effective in driving turbulence in protoplanetary disks. 
Prior studies on VSI exclusively exploit the reflecting boundary conditions (BCs) at the disk surfaces.
VSI depends critically on the boundary behaviors of waves at the disk surfaces. 
We extend earlier studies 
by performing a comprehensive numerical analysis of VSI 
with partially reflecting BCs for both the axisymmetric and non-axisymmetric unstable VSI modes. 
We find that the growth rates of the unstable modes diminish when the outgoing component of the flow is greater than the incoming one for high-order body modes. When the outgoing wave component dominates, the growth of VSI is notably suppressed. 
\deleted{Non-axisymmetric effects tend to suppress the VSI as well.}
\added{We find that the non-axisymmetric modes are unstable and they grow at a rate that decreases with the azimuthal wavenumber.} The different BCs at the lower and upper disk surfaces naturally lead to non-symmetric 
modes relative to the disk midplane. 
The potential implications of our studies for further understanding planetary formation and evolution in protoplanetary disks (PPDs) are also briefly discussed.
\end{abstract}




\section{Introduction}

Protoplanetary disks serve as the nurseries for planet formation. Turbulence is a key factor driving angular momentum transport within the disks, and further exploration of the physical factors that drive turbulence would enhance our understanding on disk mass and angular momentum transfer, which is of great significance in the processes of planet formation and evolution \citep{2007Icar..192..588Y, 2023ASPC..534..465L}. With the development of observational technology, the finer structures of protoplanetary disks are gradually emerging. The high-resolution observations by the Atacama Large Millimeter/submillimeter Array (ALMA) reveal the distribution of dust and gas within protoplanetary disks, as well as clear substructures such as rings and gaps \citep{2016A&A...592A..49T, 2024PASJ...76..437Y}. Concurrently, ALMA has detected a certain level of turbulence within protoplanetary disks, suggesting that turbulence may play a significant role in various stages of planet formation \citep{manger_vortex_2018,latter_vortices_2018, 2019ApJ...872...92F, 2024AAS...24310604F}. 
The investigation into the origin of disk turbulence will provide an indispensable theoretical framework for a more profound comprehension of the formation and evolution processes of exoplanets.

In different regions of the protoplanetary disk, the dominant mechanisms for generating turbulence are diverse \citep{2019ApJ...871..150P}. In the inner disk region, with radii smaller than about 0.3 astronomical units and the outermost disk region, beyond 100 astronomical units, turbulence is primarily driven by magnetorotational instability (MRI). In the intermediate region, however, the weak ionization results in a poor coupling between the magnetic field and gas, suppressing the MRI \citep{2011ApJ...739...51B,cb21}. Within this "dead zone", purely hydrodynamic instabilities become the primary driver of turbulence \citep{yellin-bergovoy_minimal_2021}, 
{such as the VSI \citep{urpin_magnetic_1998, urpin_comparison_2003}, zombie vortex instability (ZVI) \citep{2015ApJ...808...87M}, and convective overstability (COS) \citep{2014ApJ...788...21K}. The operation of these instabilities depends on specific environmental conditions.} VSI, due to its extensive influence in protoplanetary disks, has received widespread attention.

The VSI arises in the presence of angular velocity shear $\partial\Omega/\partial{z}$ within protoplanetary disks, serving as an effective source of turbulence. The VSI is similar in form to the Goldreich-Schubert-Fricke instability (GSF), which was initially discovered in slowly rotating stars \citep{1967ApJ...150..571G}. It does not require strong radial stratification and is induced solely by vertical shear \citep{urpin_comparison_2003}. 

For protoplanetary disks, the dependence of $\Omega$ on $z$ significantly alters the stability characteristics of the disk. However, fluid elements also experience stabilizing effects due to vertical buoyancy, hindering the growth of VSI. Effective rapid cooling can overcome this stabilizing effect caused by buoyancy. Under rapid cooling with $t_{\rm c}<\Omega_{\rm K}^{-1}h|q|/(\gamma-1)$, VSI will be effective \citep{lin_cooling_2015}, where $t_{\rm c}$ is the cooling timescale, $h$ is the disk aspect ratio, $q$ is the power-law index of temperature profile and $\gamma$ is the adiabatic index. 
{For cooling times longer than the critical cooling time of VSI, COS modes will emerge,  experiencing similar growth rates as GSF \citep{2024arXiv240415933K}.} Realistic protoplanetary disks have complex density and temperature structures, with thermal relaxation depending on the intricate spatial distribution of local disk stratification. Three-dimensional simulations have studied the thermal coupling of gas and dust particles in optically thick and optically thin regions within protoplanetary disks. The decoupling of oscillations of dust and gas particles in the upper atmosphere leads to ineffective thermal relaxation, thereby suppressing VSI turbulence \citep{pfeil_sandwich_2021}.

The VSI is classified into surface modes and body modes \citep{nelson_linear_2013,lin_cooling_2015}. 
Surface modes concentrate on the disk surface. Body modes extend vertically within the disk, exhibiting prominent vertical oscillations with radial wavenumbers much larger than vertical wavenumbers. Based on the symmetry of body modes, they can be further classified into breathing modes and corrugation modes \citep{nelson_linear_2013}.

\citet{nelson_linear_2013} studied the nonlinear evolution of VSI through numerical simulations. They showed that, although surface modes grow rapidly, unstable modes are eventually dominated by body modes, especially corrugation modes, which are characterized by {a} narrow band of large-scale vertical motions \citep{nelson_linear_2013}. This can lead to turbulence with an effective parameter $\alpha\approx10^{-4}\sim10^{-3}$, causing non-negligible angular momentum transport \citep{nelson_linear_2013,stoll_vertical_2014}.

The VSI mainly occurs in the "dead zone" with weak magnetic coupling within protoplanetary disks. The magnetic field provides a stabilizing effect through magnetic tension or magnetorotational turbulence, suppressing the VSI. With increasing magnetization, surface modes will disappear before body modes \citep{cui_vertical_2021-1,latter_vertical_2022}. The Ohmic resistivity in non-ideal MHD can overcome magnetic stability, restoring the instability \citep{cui_vertical_2021-1}. Disk winds could coexist and interact with the VSI, enhancing the accretion rate of disk winds and triggering strong VSI turbulence. Weak ambipolar diffusion strength or enhanced coupling between gas and magnetic fields will also suppress shear instability \citep{cui_global_2020}.

Previous studies on the VSI adopted a vertically global treatment with the reflecting BCs \citep[e.g.][]{nelson_linear_2013,lin_cooling_2015,cui_vertical_2021-1,2022MNRAS.514.4581S}.  
The effects of the disk's vertical extent on the VSI were studied by \citet{2015MNRAS.450...21B}. 
\cite{lin_cooling_2015} applied free and rigid BCs to the VSI. Free BCs establish a zero Lagrangian pressure perturbation at the boundary, resembling a reflecting BC with a phase difference, as we detailed in Section~\ref{appendix B}.
The rigid BC means no vertical velocity perturbation at the boundary, consistent with the one used in \citet{2015MNRAS.450...21B}.
It is also a reflecting BC, albeit with half-wave loss, as we explained in Section~\ref{sec:4.2}. This work investigates the VSI with partially reflecting BCs, introducing a new factor worth further exploration to enhance our understanding of the VSI's behavior in more realistic astrophysical scenarios.

The complex environment of protoplanetary disks at their surfaces can indeed give rise to 
the outgoing flows. 
{\citet{2024ApJ...968...29Z} considered that full disks and transition disks with small cavities have a superheated atmosphere and cool midplane, with this temperature structure producing an outgoing flow layer at optical depth $\tau_*< 1$ on top of an ingoing flow layer at $\tau_*\sim1$. } Notably, magnetohydrodynamic (MHD) disk winds are one such phenomenon that is effectively launched and driven from the disk surface, providing a physical foundation for understanding and justifying the existence of 
{outgoing} BCs. In addition, the presence of anti-symmetric magnetic field modes in protoplanetary disks results in one-sided characteristics for disk winds, meaning that these winds occur solely on one side of the disk relative to its midplane, which can potentially lead to asymmetry in the
{outgoing flow} \citep{2023arXiv231101636W}.

Consequently, it is essential to fill the research gap regarding the behavior of VSI under conditions of partial reflection at the boundaries. We focus on the effect of partially reflecting BCs, considering the axisymmetric case and further including a nonzero azimuthal wavenumber to account for non-axisymmetric VSI effects. 

The structure of the paper is as follows: In $\S2$,  we present the dynamical equations. In $\S3$, we introduce the vertical shear of the equilibrium disk model. In $\S4$, we show the linearized equations for both axisymmetric and non-axisymmetric perturbations and describe the partially reflecting BCs. In $\S5$, we analyze the growth rates and symmetry of VSI. Finally, we summarize our results in $\S6$.

\section{Basic equations}
{The} continuity, momentum and energy conservation equations of hydrodynamics for a protoplanetary disk 
{are} as follows:
\begin{equation}    
    \frac{\partial\rho}{\partial{t}}+\nabla\cdot\left(\rho\boldsymbol{u}\right)=0\ ,
    \label{eq:continuty}
\end{equation}
\begin{equation}
	\frac{\partial\boldsymbol{u}}{\partial{t}}+\left(\boldsymbol{u}\cdot\nabla\right)\boldsymbol{u}
	=-\frac{1}{\rho}\nabla{P}-\nabla\Phi\ ,
	\label{eq:momentum}
\end{equation}
\begin{equation}
	\frac{\partial{P}}{\partial{t}}+\left(\boldsymbol{u}\cdot\nabla\right){P}
	=-\gamma{P}\nabla\cdot\boldsymbol{u}-\Lambda\ ,
	\label{eq:energy}
\end{equation}
where $\rho,\boldsymbol{u}, P$ are the density, three-dimensional velocity and pressure, respectively. {To study the protoplanetary disk, we adopt the cylindrical coordinates($r,\phi,z$).} 
{The gravitational potential by the central star with mass $M$ is $\Phi=-GM/(r^2+z^2)^{1/2}$, where  $G$ is the gravitational constant. In equation ~(\ref{eq:energy}), $\gamma$ is the adiabatic index, and the sink term $\Lambda$ on the right side represents the radiative cooling.}

\section{Equilibrium disk model}
The radial power-law distribution for both density and temperature in  the cylindrical coordinate are:
\begin{equation}
	\rho(r)={\rho_0} \left( \frac{r}{r_0}\right)^p   , \ 
	T(r)=T_0\left( \frac{r}{r_0}\right)^q\ ,
\end{equation}
where $p$ is the power-law index of the mass density and $q$ is the power-low index of the temperature, $\rho_0$ and $T_0$ are the density and temperature at the fiducial radius $r_0$, respectively.  
We adopt the ideal equation of state, $P=\frac{\mathcal{R}}{\mu}\rho{T}$. Here $\mathcal{R}$ is the gas constant and $\mu$ is the mean molecular mass of gas. For a characteristic timescale $t_{\rm c}$, the radiative cooling takes the form:
\begin{equation}
	\Lambda=\frac{\rho\mathcal{R}}{\mu}\frac{T-T_{\rm{eq}}}{t_{\rm c}}
        =\frac{P-c_{\rm{s}}^2\rho_{\rm eq}}{t_{\rm c}}\ , 
\end{equation}
where $T_{\rm eq}$ and $\rho_{\rm eq}$ are the equilibrium temperature and density, respectively.  In the absence of cooling, the disk will be stable against the Solberg-Hoiland criteria. 
Sufficiently rapid cooling is an indispensable factor to trigger VSI 
since it requires rapid thermal relaxation to inhibit buoyancy.

The equilibrium state of the disk at radial and vertical directions meet:
\begin{equation}
    \frac{\partial\Phi}{\partial{r}}+\frac{1}{\rho}\frac{\partial{P}}{\partial{r}}=r\Omega^2 \ , \ 
    \frac{\partial\Phi}{\partial{z}}+\frac{1}{\rho}\frac{\partial{P}}{\partial{z}}=0 \ , \ 
    \label{eq:equbliumv}
\end{equation}
where $\Omega=\Omega(r,z)$ is the equilibrium rotation field. The density and pressure are related by the isothermal sound speed $c_{\rm s}$, which is 
{$c_{\rm s}(r)=\sqrt{ P/\rho}$}.  The density profile is: 
\begin{equation}
	\rho(r,z)=\rho(r) \exp\left[\frac{1}{{c_{\rm s}}^2}\left( \frac{GM}{\sqrt{r^2+z^2}}-\frac{GM}{r}\right) \right]\ ,
	\label{eq:density}
\end{equation}
The equilibrium angular velocity  is 
\begin{equation}
	\Omega^2(r,z)=\Omega_{\rm K}^2(r)\left[1+q+\left(p+q\right)h^2-q\frac{r}{\sqrt{r^2+z^2}}\right]\ ,
\end{equation}
where $\Omega_{\rm K}=\sqrt{GM/r^3}$ is the Kepler frequency. We assume thin disk models with a small disc aspect ratio $h$, i.e. $h=H/r\ll1$, where 
{$H=c_s/\Omega_K$} is the scale height of the disc. 
In this case, the density profile in equation~(\ref{eq:density}) is simplified as:
\begin{equation}
    \rho(r,z)=\rho(r)\exp\left(-\frac{z^2}{2H^2}\right)\ .
\end{equation}
Radial variations in temperature inevitably lead to a vertical gradient in the angular velocity, or vertical shear:
\begin{equation}
	\frac{\partial{\Omega^2}}{\partial{z}}=
        \frac{qz}{r^2}\frac{\Omega_{\rm K}^2}{\left(1+z^2/r^2\right)^{3/2}}\ .
\end{equation}
The radial gradient of the angular velocity squared is:
\begin{equation}
	\frac{\partial{\Omega^2}}{\partial{r}}=\frac{-3\Omega^2}{r}
        -\frac{qz^2}{r^3}\frac{\Omega_{\rm K}^2}{\left(1+z^2/r^2\right)^{3/2}}\ .
    \label{eq: radial gradient}
\end{equation}

\section{Linear Perturbation Equations} 
In this work, the linear perturbations for both the axisymmetric and non-axisymmetric cases with partially reflecting BCs will be discussed. 
The Eulerian perturbations ($\delta u_r,\delta u_{\phi},\delta u_z,\delta\rho,\delta{P}$) have space and time dependence of the form 
{$\exp(i{k_r}r+im\phi-i\upsilon{t})$}, where $k_r$ is the radial wavenumber, $m$ is the azimuthal wavenumber and $\upsilon$ is the complex frequency. The linearized perturbation equations for the velocity, density and pressure perturbations are:
\begin{equation}
	i\left(\upsilon-m\Omega\right)\frac{\delta\rho}{\rho}
	=i{k_r}\delta{u_r}{+\frac{\delta{u_r}}{r}}
	+\frac{im}{r}\delta{u_{\phi}}
	+\frac{\partial\delta{u_z}}{\partial{z}}
	+\frac{\partial{\ln}\rho}{\partial{z}}\delta{u_z}  \  ,
	\label{eq:perturbation 1}
\end{equation}
\begin{equation}
	i\left(\upsilon-m\Omega\right) \delta{u_r}=i{k_r}\frac{\delta P}{\rho}-2\Omega\delta{u_{\phi}} \ ,
	\label{eq:perturbation 2}
\end{equation}
\begin{equation}
	i\left(\upsilon-m\Omega\right) \delta{u_{\phi}}
	=\frac{im}{r}\frac{\delta P}{\rho} + r\frac{\partial{\Omega}}{\partial{z}}\delta{u_z}
	+\frac{\kappa^2}{2\Omega}\delta{u_r} \ ,
	\label{eq:perturbation 3}
\end{equation}
\begin{equation}
	i\left(\upsilon-m\Omega\right) \delta{u_z}=\frac{d}{dz}\left(\frac{\delta P}{\rho}\right)+\frac{\partial{\ln\rho}}{\partial{z}}\left(\frac{\delta P}{\rho}-{c_{\rm s}}^2\frac{\delta\rho}{\rho}\right) \  ,
	\label{eq:perturbation 4}
\end{equation}
\begin{align}
		i\left(\upsilon-m\Omega\right) \frac{\delta P}{\rho}&={c_{\rm s}}^2{\gamma}\left(i{k_r}\delta{u_r}{+\frac{\delta{u_r}}{r}}+\frac{im}{r}\delta{u_{\phi}}+\frac{\partial{\delta{u_z}}}{\partial{z}}\right)\nonumber\\
		&\qquad\quad+{c_{\rm s}}^2\frac{\partial{\ln\rho}}{\partial{z}}\delta{u_z}
		+\frac{1}{t_{\rm c}}\left(\frac{\delta P}{\rho}-{c_{\rm s}}^2\frac{\delta\rho}{\rho}\right) \  ,
		\label{eq:perturbation 5}
\end{align}
where $\kappa$ is the epicyclic frequency and   {$\kappa^2\equiv4\Omega^2+r\frac{\partial\Omega^2}{\partial{r}}$}. 
{Note that our equations retain the curvature terms neglected in \citet{lin_cooling_2015}. The impact caused by the curvature terms is weak.
}  
We have ignored the radial gradients of the equilibrium density and pressure to simplify the equation. {This type of radially local approximation was initially proposed in \citet{MP14} and has also been adopted in \citet{lin_cooling_2015}.} {However, we retain the radial temperature gradient to ensure the VSI operates.}

\subsection{Wave Equations for the Axisymmetric VSI}\label{sec:4.1}
 For {the} axisymmetric VSI, we take $m=0$. In terms of the variables $y_1\equiv\delta{u_z}$, and $y_2\equiv\delta{P}/\rho$,  Equations~(\ref{eq:perturbation 1})-(\ref{eq:perturbation 5}) can be cast into two first-order ordinary differential equations (ODEs) \footnote{In a more compact matrix form, 
 \[ 
	\frac{d \boldsymbol{Y}}{d \hat{z}}  = {\boldsymbol{A}} \boldsymbol{Y} \ , \   {\rm where}  \  {\boldsymbol{A} } = \begin{pmatrix} A_{11} & A_{12} \\ A_{21} & A_{22} \end{pmatrix}  \ , \  \boldsymbol{Y} \equiv (y_1, y_2)^T = \begin{pmatrix} y_1 \\ y_2
 \end{pmatrix} \ .
\] 
It should be stressed that all the wave propagation properties are hidden in the matrix $\boldsymbol{A}$. The two eigenvalues ($i\hat{k}_{z1}$, $i \hat{k}_{z2}$) and two eigenvectors ($\boldsymbol{r}_1$, $\boldsymbol{r}_2$) of the matrix $\boldsymbol{A}$ are important for our analysis of BCs. (Detailed derivation of the eigenvalues and eigenvectors are given in Appendix~\ref{appendix B}. Further explanation can be found in Appendix~\ref{appendix A}.)}:
 \begin{equation}
     \frac{d y_1}{d \hat{z}} = A_{11} y_1 + A_{12} y_2  \  ,  \ 
     \label{eq: eigenfunc 1}
     \frac{d y_2}{d \hat{z}} = A_{21} y_1 + A_{22} y_2 \ .
 \end{equation}
 The relevant coefficients in the above linear ODEs are
 \begin{equation}
     A_{11} = (\chi + Q) \hat{z} \ , \  A_{12} = {i\hat{\upsilon}}\left(\chi+U\right) \ , 
 \end{equation}
 \begin{equation}
     A_{21} = i\hat{\upsilon}- \frac{\hat{z}^2\left(\chi-1\right)}{i\hat{\upsilon}}   \ , \ A_{22} = - \left(\chi-1\right)\hat{z} \ ,  
 \end{equation}
 where $\chi=(1-i\hat{\upsilon}\beta)/(1-i\hat{\upsilon}\beta\gamma)$, $Q = ihq\hat{k}_r+qh^2$, and $U=\hat{k}_r^2-i\hat{k}_rh$.  
 Note that the equations above {have} been written in a dimensionless form adopting the following rules: $z=\hat{z}H$, ${k_r}=\hat{k_r}/H$,  $\upsilon=\hat{\upsilon}\Omega_{\rm K}$, $\beta={t_{\rm c}}\Omega_{\rm K}$.  
 The eigenvalue $\hat{\upsilon}$ is a complex number. Its real part $\hat{\omega}$ is the frequency and the imaginary part $\hat{\sigma}$ means the growth rate. {With our convention, the sound speed $c_s=H\Omega_K$ is normalized to unity.} 

{ 
We are interested in the frequency range in which $\upsilon$ is small compared to the epicyclic frequency $\kappa$. This low-frequency approximation filters out the acoustic waves and retains the inertial-gravity waves \citep{1993ApJ...409..360L}.}
With this approximation and rapid cooling $\beta = 10^{-3}$, the dispersion relation~(\ref{eq: dispersion equation}) in Appendix~\ref{appendix A} can be written as
    \begin{equation}
    \hat{\upsilon}^2\simeq
    \frac{\hat{k}_z^2+i\left(ihq\hat{k}_r+qh^2+1\right)\hat{z}\hat{k}_z}
    {1+\hat{k}_r^2-i\hat{k}_rh} \ ,
    \end{equation}
Since $h\ll1$ and $\hat{k}_r\gg1,\hat{k}_z$, the equation shows the  dispersion relation for inertial waves {\citep{2009ApJ...696.2054G}}
\begin{equation}
    {\upsilon}^2\approx\Omega_K^2
    \frac{\hat{k}_z^2+i\left(ihq\hat{k}_r+1\right)\hat{z}\hat{k}_z}
    {\hat{k}_r^2}\sim
    \Omega_K^2\frac{\hat{k}_z^2}{\hat{k}_r^2}\ .
\end{equation}
It is worthwhile to note that the VSI actually corresponds to the inertial wave. 
The vertical component of the group velocity and phase velocity for inertial waves are always in the same direction. 

\subsection{Wave Equations for the Non-axisymmetric VSI}
In the non-axisymmetric case, the azimuthal wavenumber $m\neq0$. We still adopt the low-frequency approximation to simplify our analysis, i.e., the Doppler shifted frequency is much smaller than the epicyclic frequency, {${|\upsilon-m\Omega|}^2\ll\kappa^2$}. 
The governing ODEs for the non-axisymmetric case read:
\begin{equation}
     \frac{d y_1}{d \hat{z}} = A_{11}^\prime y_1 + A_{12}^\prime y_2   \   ,   \ 
     \label{eq: eigenfunc_m 1}
     \frac{d y_2}{d \hat{z}} = A_{21}^\prime y_1 + A_{22}^\prime y_2 \ , 
 \end{equation}
where
 \begin{equation}
     A_{11}^\prime = \left( \chi{^{\prime}} +Q^\prime \right) \hat{z} \ , \   
 \end{equation}
 \begin{equation}
     A_{12}^\prime = {i\left(\hat{\upsilon}-m\right)} \left(\chi{^{\prime}}+U^\prime\right)
   +\frac{W}{2}-\frac{\hat{k}_rmqh^3}{2}\hat{z}^2 \ ,
 \end{equation}
 \begin{equation}
     A_{21}^\prime = i\left(\hat{\upsilon}-m\right)-\frac{\hat{z}^2\left(\chi{^{\prime}}-1\right)}{i\left(\hat{\upsilon}-m\right)} \ , \ A_{22}^\prime = -\left(\chi{^{\prime}}-1\right)\hat{z} \ .
 \end{equation}
 Relevant variables in the above matrix coefficients are 
 \[
\chi^\prime = \left[1-i(\hat{\upsilon}-m)\beta\right]/\left[1-i(\hat{\upsilon}-m)\beta\gamma\right] \ , \ U^\prime=U+m^2h^2\ ,
 \]
 \[
 Q^\prime=Q+{\left(\hat{{\upsilon}}-m\right)mqh^2}/{2} \ , \ W=4imh^2-3mh\hat{k}_r \ , 
 \]
respectively. Compared to equation~(\ref{eq: eigenfunc 1}), the additional terms in the above equations are due to the non-axisymmetric effect. 

\subsection{Boundary Conditions}\label{sec:4.2}
\added{We adopt the WKB approximation to specify the BCs. Note that it is a local approximation of the linear second-order ODEs. The WKB approximation is only exploited locally at the boundary, which provides certain constraints on the global integration. Our global solutions are obtained by numerical integration rather than the WKB approximation. According to Appendix~\ref{appendix B}, the WKB approximation can be understood in terms of the eigenstructures of the coefficient matrix of the wave equations.} 
The variables $\boldsymbol{Y}$  $\equiv (y_1, y_2)^{T}$ can be viewed as the sum of two eigenvectors $\boldsymbol{r}_1$ and $\boldsymbol{r}_2$:
 \begin{equation}
    {\boldsymbol{Y} } = {a_1}{\boldsymbol{r}_1}+{a_2}{\boldsymbol{r}_2} \ .
     \label{eq:23}
 \end{equation}
As mentioned in Appendix~\ref{appendix A}\footnote{The identification of the wave propagation direction is discussed in Appendix~\ref{appendix A}.}, we know that the wave group velocity associated with ${\boldsymbol{r}_1}$ is incoming and ${\boldsymbol{r}_2}$ outgoing. The coefficients $a_1$ and $a_2$ represent the strength of waves in these two directions, respectively. 
The boundary parameter $R\equiv{a_2}/{a_1}$ is determined by $y_1$ and $y_2$ in the following way (Please refer to Appendix \ref{appendix B} for the derivation):
\begin{equation}
	R=\dfrac
	{\left[-{A_1}\hat{z}+i\sqrt{ {A_3}-{A_2}\hat{z}^2}\right] {y_1}-{B_1}{y_2}}
	{\left[{A_1}\hat{z}+i\sqrt{{A_3}-{A_2}\hat{z}^2}\right] {y_1}+{B_1}{y_2}}\ .
    \label{eq: bc}
\end{equation}
The parameters, $A_1$, $A_2$, $A_3$ and $B_1$ in equation~(\ref{eq: bc}) are explicitly  given in Appendix~\ref{appendix B}. Note that equation (\ref{eq: bc}) can be written alternatively as a linear combination of $y_1$ and $y_2$ :
\begin{equation}	
        \left({A_1}\hat{z}+\frac{R-1}{R+1}i\sqrt{{A_3}-{A_2}\hat{z}^2}\right){y_1}+{B_1}{y_2}=0 \ .
	\label{eq:boundary}
\end{equation}
The above equation is the actual BC adopted in our numerical calculations. 
Perfect reflection occurs when $R=1$, 
\begin{equation}
	{A_1}\hat{z}{y_1}+{B_1}{y_2}=0 \ . 
\end{equation}
For a reflection with a half-wave loss, the BC can be expressed as $y_1=0$ {with $R=-1$}, i.e.  $\delta{u_z}=0$ at $\hat{z}=\pm\hat{z}_{\rm max}$. 
{As $R$ approaches infinity, the mode solely exhibits an outgoing 
inertial wave. Purely outgoing boundary conditions are widely used in the studies of Rossby wave instability \citep{2000ApJ...533.1023L, 2013MNRAS.429.2748Y, 2022MNRAS.514.1733H}.}

The BC for the non-axisymmetric case could be similarly obtained, which is:
\begin{equation}
    \left({A_1^{\prime}}\hat{z}+
    \frac{R-1}{R+1}i\sqrt{{A_3}^{\prime}+{A_4^{\prime}}\hat{z}^4-{A_2}^{\prime}\hat{z}^2}\right){y_1}+\left({B_1}^{\prime}-{B_2}^{\prime}\hat{z}^2\right){y_2}=0 \ ,
    \label{eq:boundary_m}
\end{equation}
where 
\begin{equation}
    {B_1^{\prime}}=2{i\left(\hat{\upsilon}-m\right)}\left(\chi{^{\prime}}+U^\prime\right)+W  \ , \
    {B_2^{\prime}}=\hat{k}_rmqh^3 \ , 
\end{equation}
\begin{equation}
    A_1^{\prime}=2\chi^{\prime}+Q^\prime -1 \ ,
\end{equation}
\begin{equation}
    A_2^{\prime}=
    \left(Q^\prime+1\right)^2{-2i(\hat{\upsilon}-m){B_2^{\prime}}}
    +4\left(\chi{^{\prime}}-1\right)\left(Q^\prime-V\right)\ ,
\end{equation}
\begin{equation}
    A_3^{\prime}=4\left(\hat{\upsilon}-m\right)^2\left(\chi{^{\prime}}+V\right) \ , \ 
A_4^{\prime}=\frac{2i\left(\chi{^{\prime}}-1\right)}{\hat{{\upsilon}}-m}{B_2^{\prime}} \ ,
\end{equation}
respectively. The parameter $V$ is given by $V=U^\prime-iW/[2(\hat{\upsilon}-m)]$. 
{Following \citet{1993ApJ...409..360L}, it should be noted that when applying the low-frequency approximation to non-axisymmetric analyses, a critical requirement exists that the modes must be tightly wrapped, leading to a condition $m \ll \hat{k}_r/h$.}

\section{Numerical result}
To investigate the axisymmetric or non-axisymmetric linear perturbations of VSI, we need to solve Equation~(\ref{eq: eigenfunc 1}) or Equation~(\ref{eq: eigenfunc_m 1}).  
The eigenvalues of these equations are closely related to the BCs, which we adopt as Equation~(\ref{eq:boundary}) or Equation~(\ref{eq:boundary_m}). These BCs represent a linear combination of outgoing and incoming wave components of different strengths. The numerical solution of this kind of complex two-point boundary eigenvalue problem can be efficiently obtained using the relaxation method {\citep{Press1992}}. In this method, we discretize the ODEs into finite difference equations (FDEs) across a mesh that spans the domain of interest. Through iterative trial and error, we progressively determine eigenvalues and eigenfunctions by a method analogous to the Newton$-$Raphson method. Throughout these iterations, both the ODEs and BCs are satisfied simultaneously. In this study, we employ 5001 uniform mesh points to achieve a high-accuracy numerical solution with an average error of less than $10^{-8}$. The basic parameters of the disk are set as $q=-1$, $p=-1.5$, $h=0.05$, $\gamma=1.4$, respectively.



\begin{figure}
    \centering
    \includegraphics[width=\columnwidth]{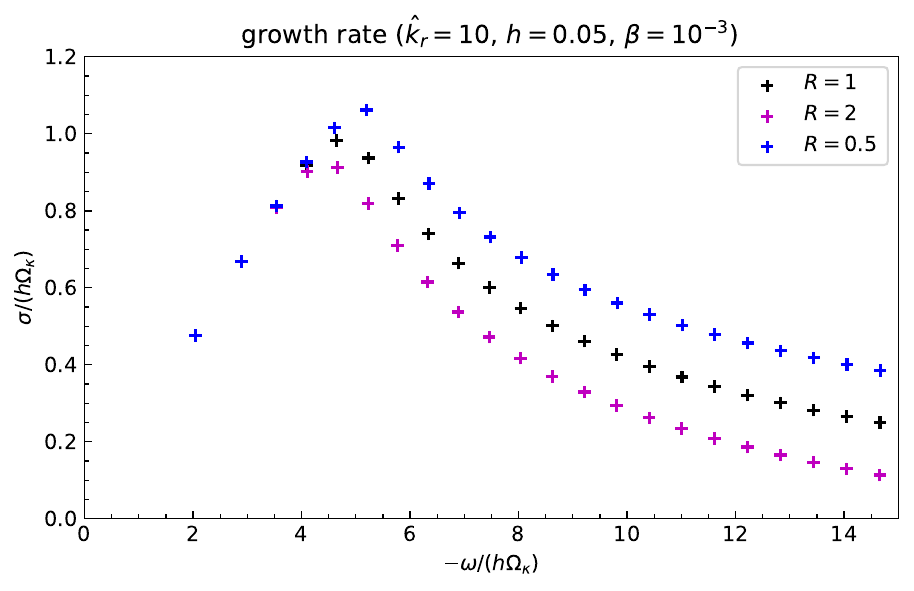}
    \includegraphics[width=\columnwidth]{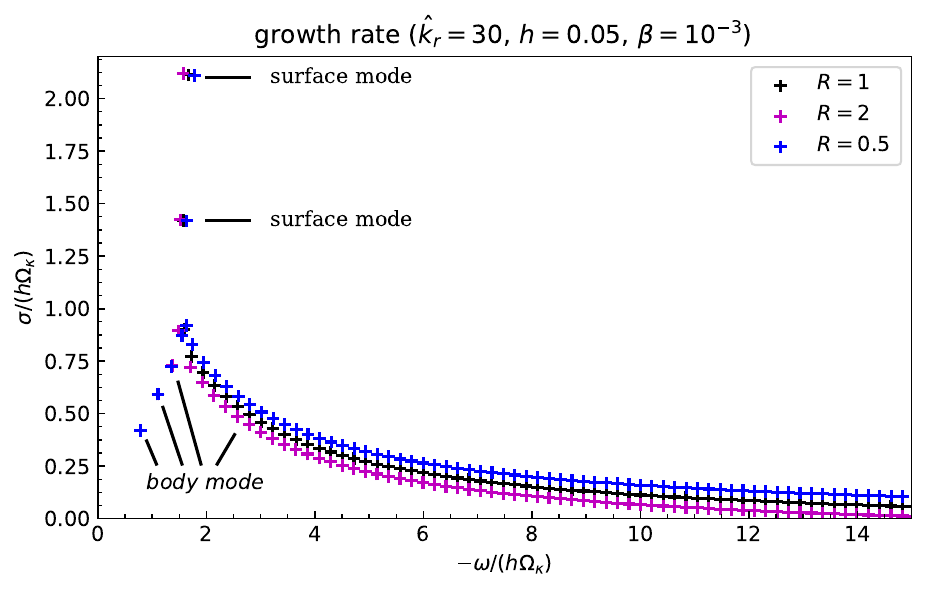}
    \caption{Variations of the growth rate with frequency for the radial wavenumber ${\hat{k}_r}=10$ and ${\hat{k}_r}=30$, respectively. The cooling parameter is $\beta=10^{-3}$. The disk respect ratio $h=0.05$ and maximum disk height $z_{\rm max}=5H$. The boundary condition parameter is set to $R=2$ (magenta pluses), $R=0.5$ (blue pluses) and $R=1$ (black pluses), respectively.  
    }
    \label{fig:growthrate b-3}
\end{figure}
\subsection{Axisymmetric VSI unstable modes}
\begin{figure}
    \centering
    \includegraphics[width=\columnwidth]{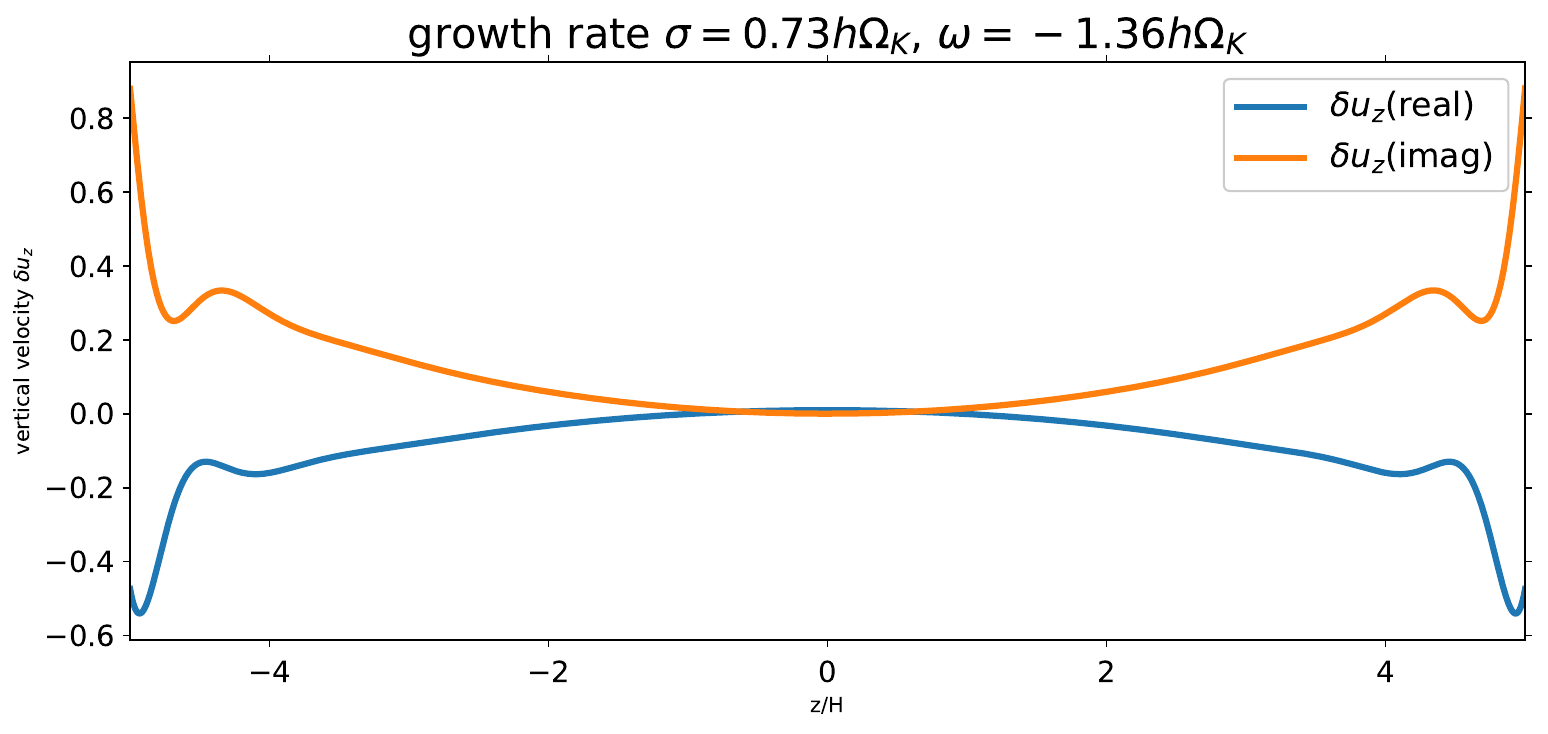}
    \includegraphics[width=\columnwidth]{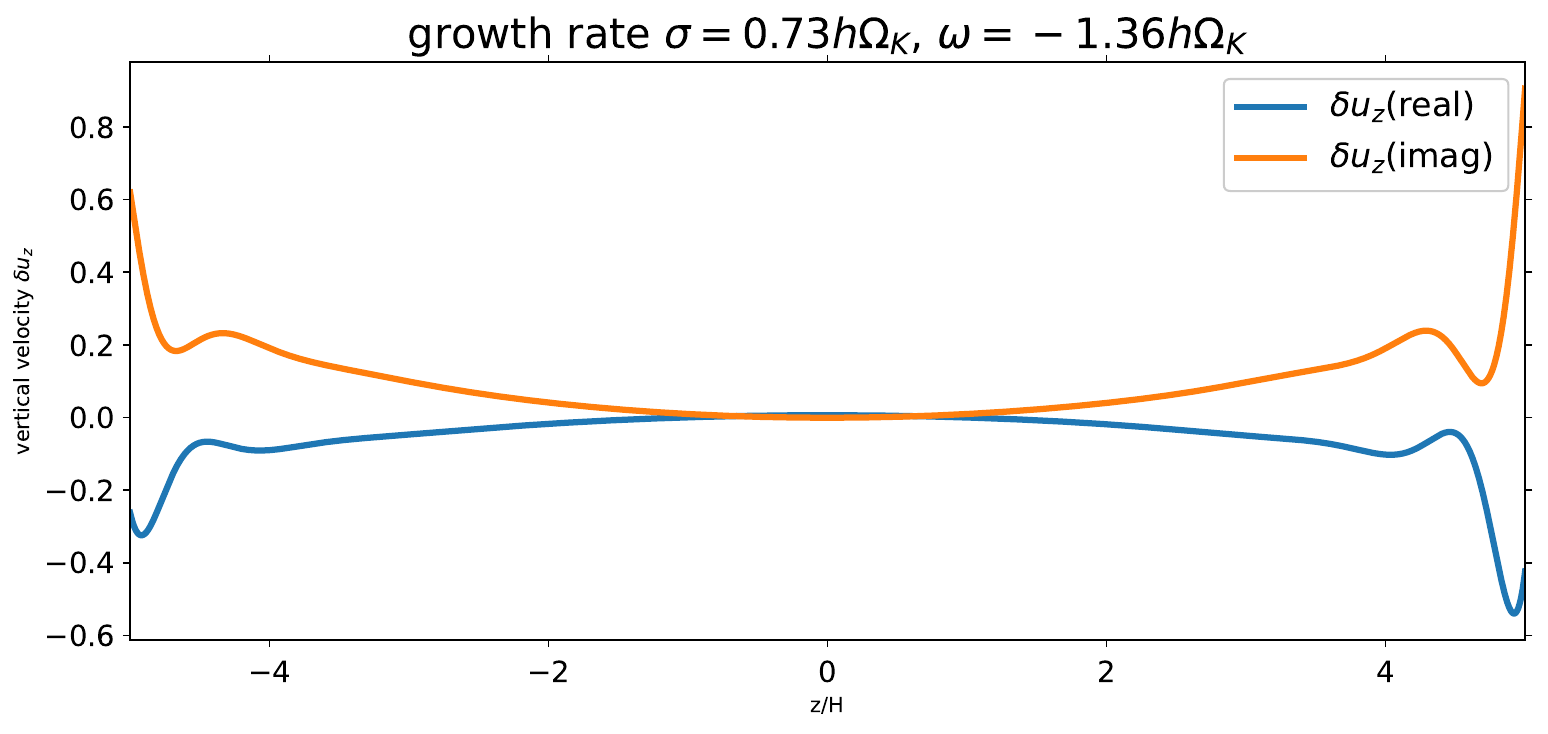}
    \caption{Vertical velocity perturbation $\delta{u_z}$ of the low order corrugation mode. Top: with same boundary parameter $R=1$ at $z=\pm z_{\rm max}$. Bottom: with boundary parameter $R=1$ on the lower boundary and boundary parameter $R=2$ on the upper boundary, respectively. }
    \label{fig:eigenfunc}
\end{figure}

\begin{figure}
    \centering
    \includegraphics[width=\columnwidth]{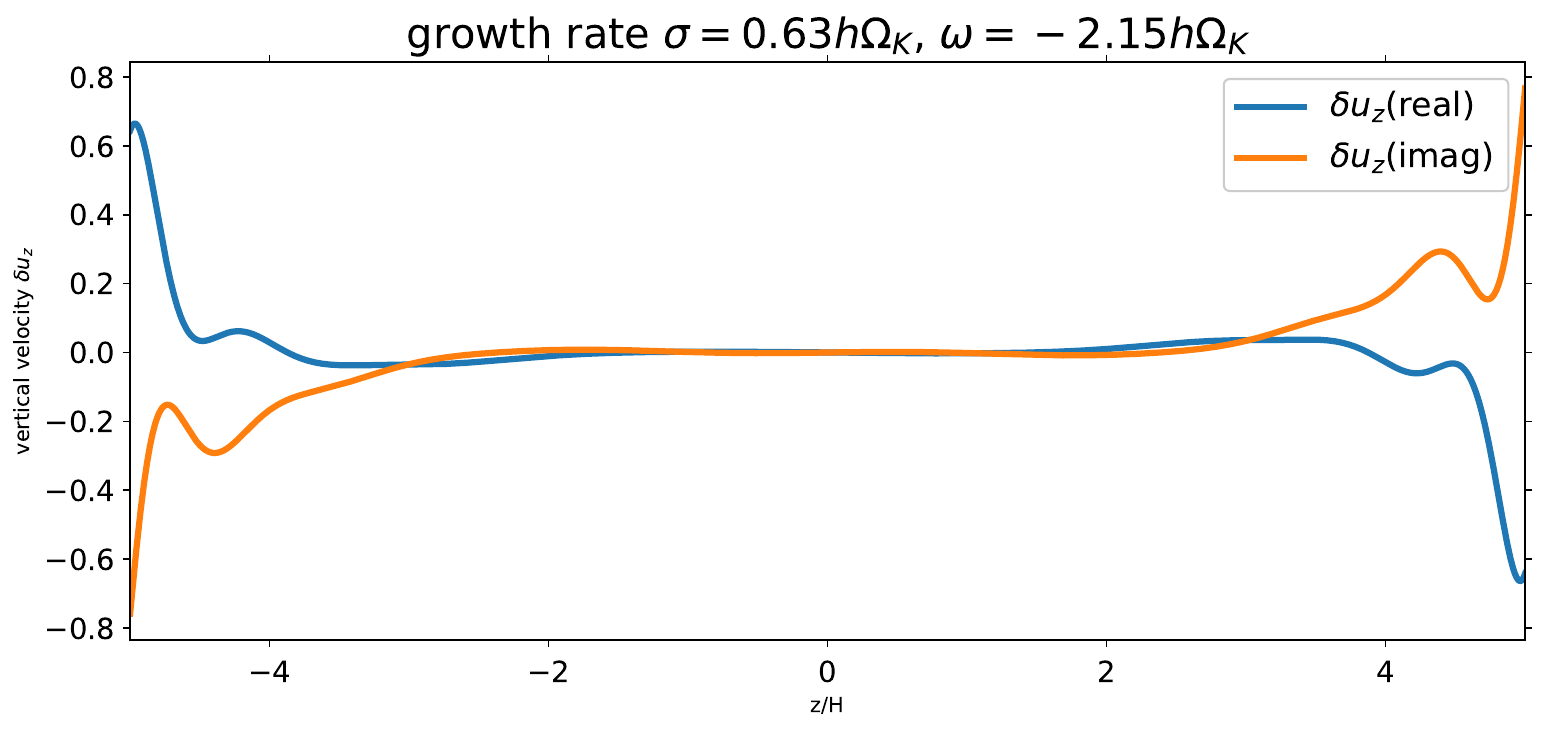}
    \includegraphics[width=\columnwidth]{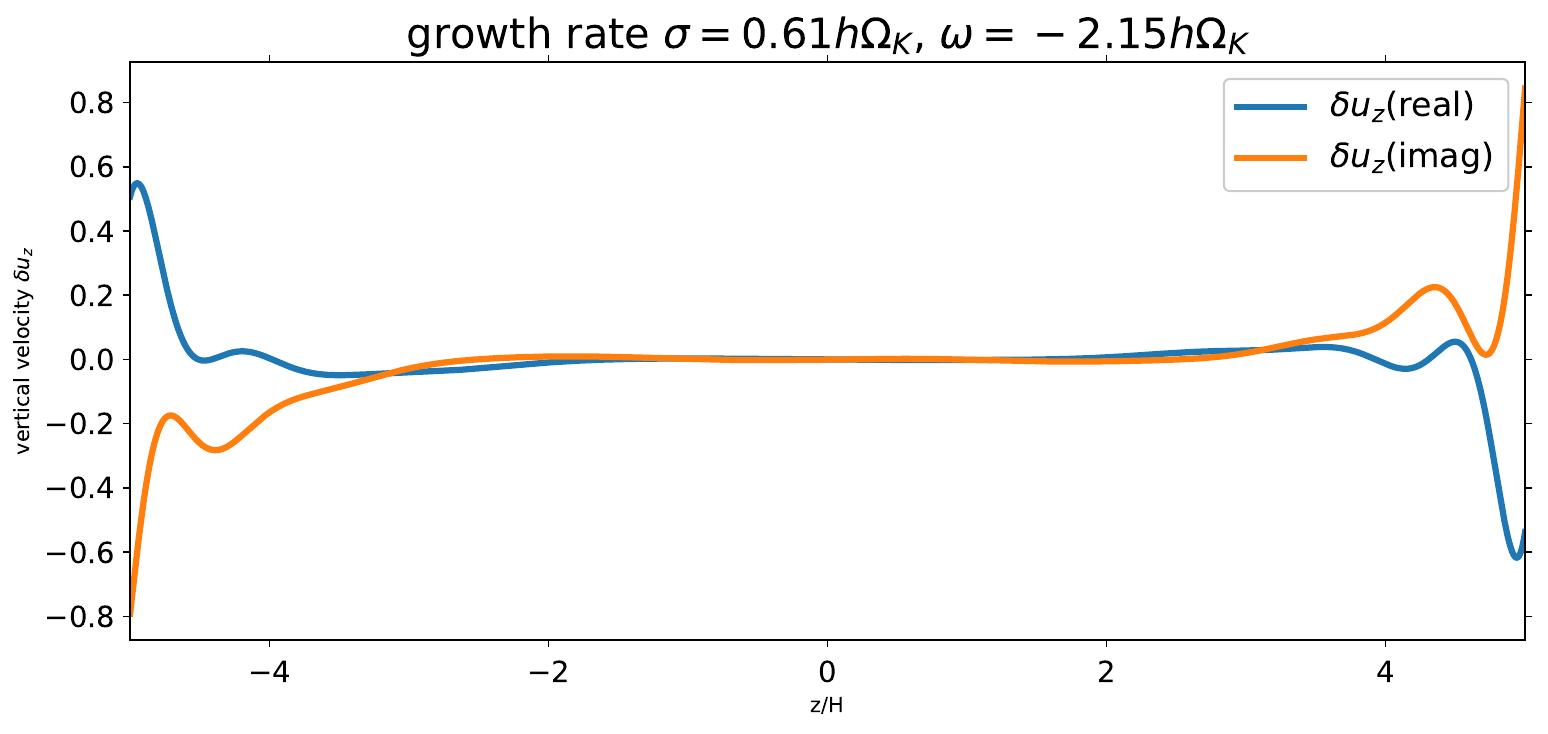}
    \caption{Same as Figure~\ref{fig:eigenfunc} but for higher order breathing mode.}
    \label{fig:eigenfunc_1}
\end{figure}

{Usually, VSI modes exhibit odd or even symmetry. Our calculations shown in Fig.~\ref{fig:eigenfunc}-\ref{fig:eigenfunc_1} with symmetric BCs at $z=-z_{\rm max}$ and $z=z_{\rm max}$ obtain similar results to \citet{nelson_linear_2013}. However, there exist highly asymmetric cases, as demonstrated in the bottom right panel of Figure~20 in \citet{nelson_linear_2013}. There is still a lack of a clear explanation for the asymmetry. For simplicity, we choose the domain $z\in [0,z_{\rm max}]$  to avoid such asymmetric surface modes.
}
At the midplane $z=0$, we set the BCs according to the parity of the eigenfunction. While at $z=z_{\rm max}$, we set partially reflecting BCs. 

Our numerical results with the cooling parameter $\beta=10^{-3}$ are shown in Fig.~\ref{fig:growthrate b-3}, which depicts the variation of growth rate with the frequency for partially reflecting BCs.
Each symbol denotes a specific order mode. For a general set of parameters, lower-order modes possess a smaller eigenfrequency $|\omega|$.
In the upper panel, we present the result for smaller radial wavenumber ${\hat{k}_r} = 10$. Only the body modes are observed in this case. The results for larger radial wavenumber ${\hat{k}_r}=30$ are displayed in the lower panel. In this case, both the body and surface modes show up. 
The growth rate varies with the frequency in a non-monotonic way. With the gradual increase of frequency, the growth rate first increases with the frequency (i.e., lower-order modes). Once the growth rate reaches a maximum, it would decrease with the frequency (i.e., higher-order modes). 
The growth rate of the low order body mode is almost not affected by the BCs. However, for the higher-order body modes, the influence of the BCs becomes more obvious. However, for surface modes, the boundary has a relatively minor effect on the growth rate.
As the parameter $R$ decreases, the growth rates become larger. On the other hand, as the parameter $R$ increases, the growth rates decrease. Our result indicate that the surface of the disk actually acts as a trap of the inertial waves. When the trapping effects become weaker ($R$ becomes larger), the VSI would become weaker as well.


Fig.~\ref{fig:eigenfunc} shows the vertical velocity perturbation, $\delta{u_z}$, of the low-order corrugation mode (symmetric with respect to midplane) with growth rate $\sigma=0.73h\Omega_{\rm K}$.
In the upper panel, perfectly reflecting BC is set at both the upper and lower boundaries. The resulting eigenfunction is symmetric relative to the midplane. In the lower panel, the perfectly reflecting BC ($R=1$) is adopted at the lower boundary $z=-5H$ and a partially reflecting BC ($R=2$) at the upper boundary $z=5H$. 
The strength of the 
{outgoing wave} is twice that of the 
{incoming} wave, breaking the symmetry with respect to the midplane. At the upper boundary, where the flow is predominantly outgoing, the vertical velocity perturbations are stronger than those observed at the lower boundary.
Owing to its nature as a low-order mode, the growth rate is barely affected by BCs. 
For higher order modes, such as the breathing mode (anti-symmetric with respect to the midplane) {with a frequency $|\omega|=2.15h\Omega_K$} shown in Fig.~\ref{fig:eigenfunc_1}, the growth rate is clearly affected by the BCs.
For the partially reflecting BC with $R=2$, the growth rate has been suppressed by $0.02h\Omega_K$. Meanwhile, the anti-symmetric configuration originally relative to the midplane can no longer be maintained. 

\begin{figure}
    \centering
    \includegraphics[width=\columnwidth]{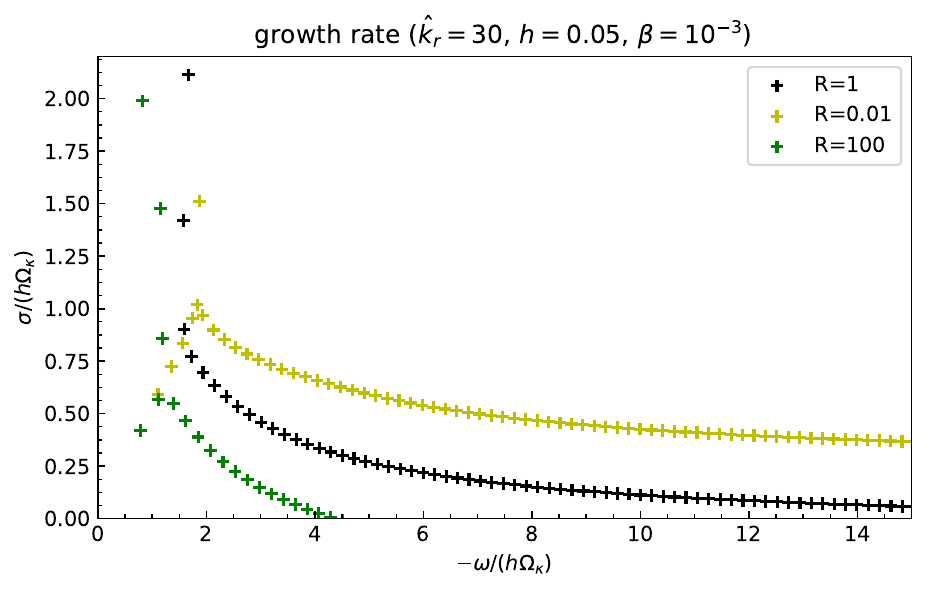}
    \caption{Growth rate and intrinsic frequency of the unstable modes for $\hat{k}_r=30$ and extreme boundary parameters $R=0.01$ (yellow pluses), $R=100$ (green pluses) and the case of $R=1$ (black pluses) as a reference.}
    \label{fig:growthrate Req100}
\end{figure}

We compare three cases of $R=0.01$, $R=1$, and $R=100$ in Fig.~\ref{fig:growthrate Req100}. When the strength of the 
{outgoing wave} is much greater than that of the 
{incoming wave}, the growth rate is significantly suppressed. Especially, the growth rate of the higher order modes with frequency 
{$|\omega|\gtrsim 5h\Omega_{\rm K}$} are preferentially damped. An increase in the strength of 
{the outgoing component} also results in a shift of the wave frequency where the maximum growth rate is achieved towards the lower frequency. We find that the VSI is quite similar to the Papaloizou-Pringle instability (PPI), which requires a reflecting boundary for PPI to operate~\citep{1984MNRAS.208..721P}. As a comparison, there exists more robust instability, such as Rossby wave instability \citep{1999ApJ...513..805L, 2022MNRAS.514.1733H}. The reflecting boundary is not necessary for RWI. The fully outgoing boundary conditions (i.e., $R=\infty$) are often adopted in the RWI calculations. 

\begin{figure}
    \centering
    \includegraphics[width=\columnwidth]{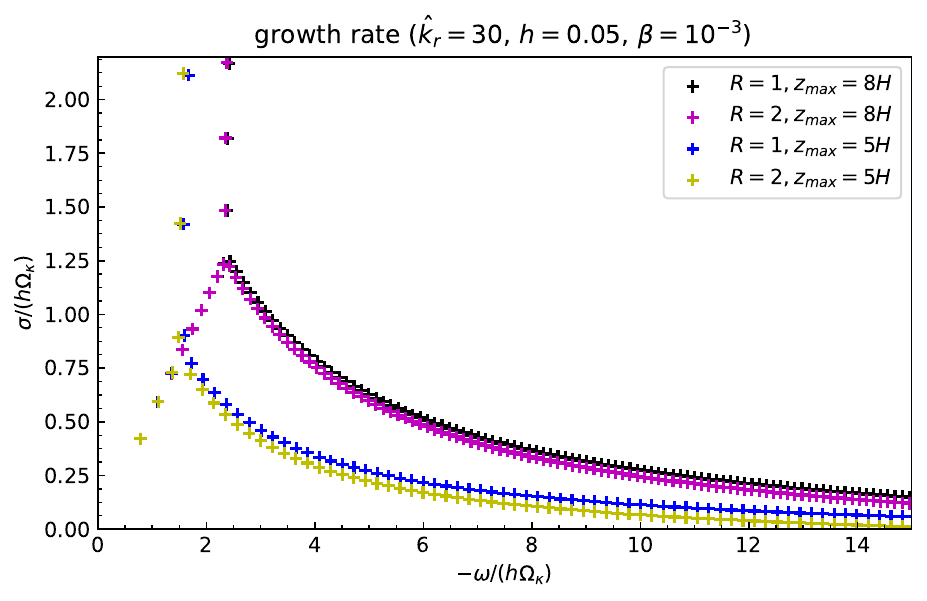}
    \caption{Growth rate of unstable modes vs. vertical extent. Blue and yellow pluses symbolize boundary parameters $R=1$ and $R=2$ for $z_{\rm max}=5H$, while black and magenta pluses represent the same parameters for $z_{\rm max}=8H$.}
    \label{fig:growthrate zmax}
\end{figure}
 
In Fig. {\ref{fig:growthrate zmax}}, we note that the growth rate varies non-monotonically with the frequency. Below a certain threshold frequency, the growth rate increases with the frequency. Above the threshold, the growth rate decreases with the frequency.
The disk vertical extent has an important effect on the threshold frequency and corresponding growth rate.
This phenomenon applies not only to reflecting BCs but also to partially reflecting BCs, as shown in Fig.~\ref{fig:growthrate zmax}.
The threshold frequency for $z_{\rm max} = 8 H$ is larger than the threshold for $z_{\rm max} = 5 H$.
A larger vertical extent has a higher growth rate at the threshold frequency as well.
Comparing the changes in growth rates due to contrasting BCs across the two vertical extents, the partially reflecting BCs have less effect on the growth rate in the larger one. {Even so, as the vertical extent increases, a very high $R$ still significantly suppresses the unstable modes, similar to the findings in Fig.~\ref{fig:growthrate Req100}}.

\begin{figure}
    \centering
    \includegraphics[width=\columnwidth]{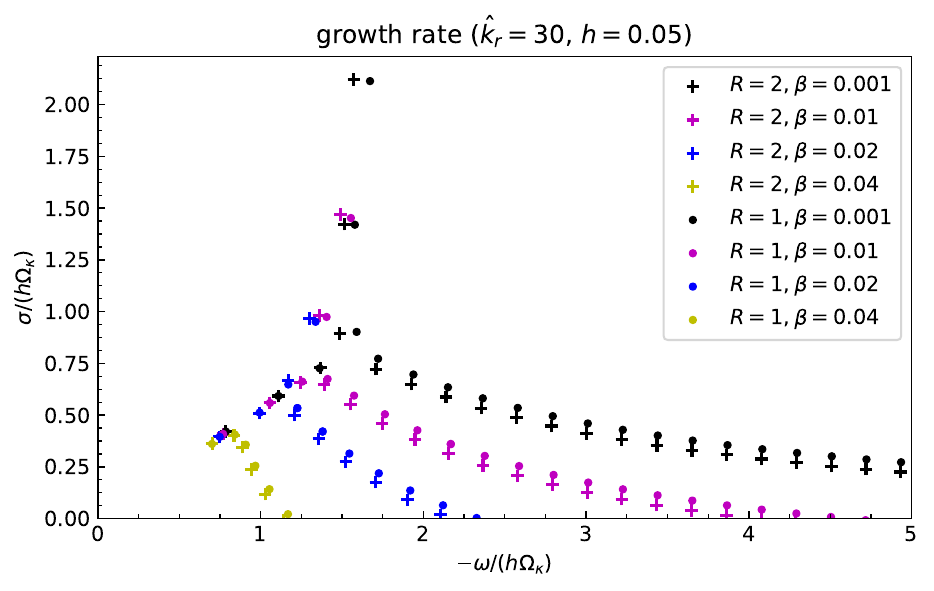}
    \caption{Growth rate and wave frequency of the unstable modes with $\hat{k}_r=30$, disk respect ratio $h=0.05$, 
    and thermal relaxation timescale $\beta=0.001$ (black), $\beta=0.01$ (magenta), $\beta=0.02$ (blue), $\beta=0.04$ (yellow) {for different boundary parameters $R=2$ (colored pluses) and $R=1$ (colored dots).}}
    \label{fig:growthrate cooling}
\end{figure}

Since the cooling is essential for VSI to work, it would be interesting to study how the cooling affects VSI with our new BCs.  
We compare in Fig.~\ref{fig:growthrate cooling} the mode growth rate and frequency for different values of cooling parameter $\beta$ with $R=2$ (partially reflecting) and $R=1$ (perfectly reflecting).
It shows that, for either partially reflecting or perfectly reflecting BCs, the growth rates increase as $\beta$ varies from 0.04 to 0.001. At sluggish thermal relaxation, the growth of unstable modes is notably reduced. In this case, only the lower order unstable modes can survive. 
Furthermore, even with slower cooling, VSI is still affected by BCs.
The augmentation of outgoing wave leads to a reduction in the growth rate of higher order modes, consistent with the scenario observed when cooling is set at 0.001 (as illustrated in Fig.~\ref{fig:growthrate b-3}).

\subsection{Non-axisymmetric VSI unstable modes}

\begin{figure}
    \centering
    \includegraphics[width=\columnwidth]{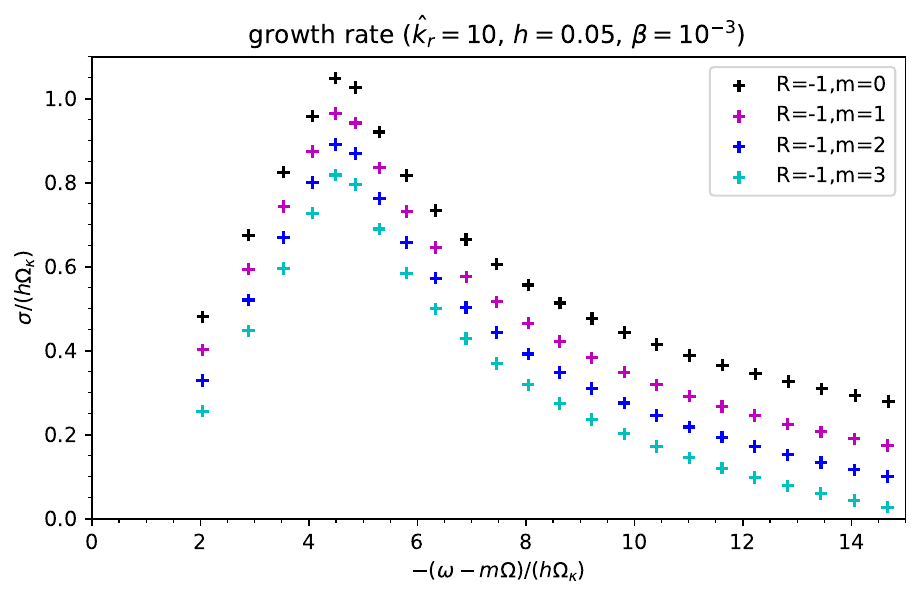}
    \caption{{Growth rate and frequency of the unstable modes with radial wavenumber $\hat{k}_r=10$, disk aspect ratio $h=0.05$, cooling time $\beta=10^{-3}$. We set perfectly reflecting BCs $R=-1$ and azimuthal wavenumbers $m=0$ (black pluses), $m=1$ (magenta pluses), $m=2$ (blue pluses), $m=3$ (cyan pluses).}}
    \label{fig:non-axisy k10}
\end{figure}
{ In Fig.~\ref{fig:non-axisy k10}, we show our results of non-axisymmetric VSI. The results for $\hat{k}_r=10$ with both $m\neq0$ (non-axisymmetric) and $m=0$ (axisymmetric) are displayed. The parameter $R=-1$ means perfect reflection with a half-wave loss. With the increase of $m$, there is an overall decrease in the growth rate. 
Note that the azimuthal wavenumber $m\ll\hat{k}_r/h$, thus the low-frequency approximation could be used to deal with the non-axisymmetric perturbations.}

\begin{figure}
    \centering
    \includegraphics[width=\columnwidth]{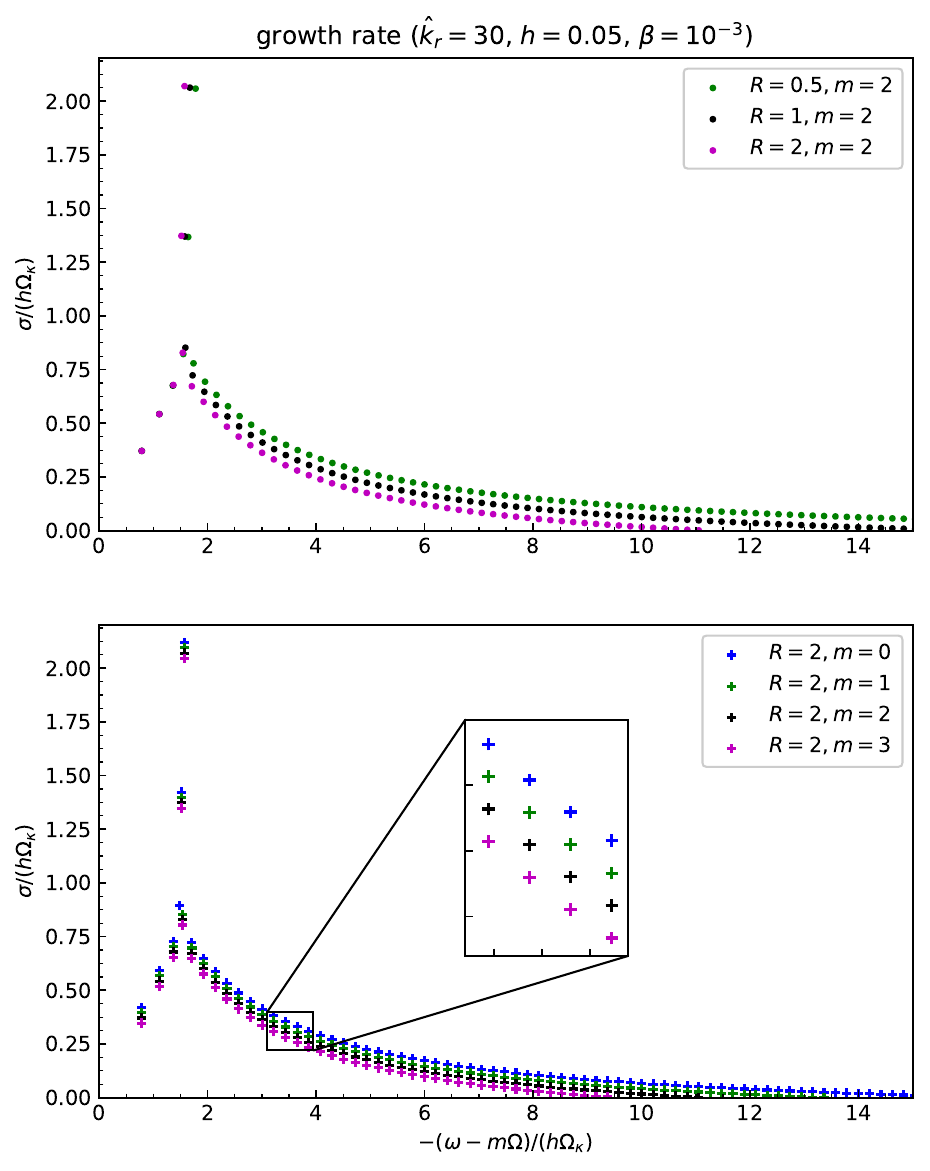}
    \caption{{unstable modes with different BCs and same azimuthal wavenumber (top), same BC but different azimuthal wavenumbers (bottom).}}
    \label{fig:non-axisymmetric}
\end{figure}

{Fig.~\ref{fig:non-axisymmetric} shows non-axisymmetric VSI with larger radial wavenumber for both the body and surface modes. The top panel shows the effect of BCs on non-axisymmetric unstable modes with $m=2$. 
Consistent with the trend in the axisymmetric scenario, boundaries characterized by a stronger outgoing component suppress the growth rate of non-axisymmetric instability as well. 
The bottom panel elucidates the variation in the growth rate caused by non-axisymmetric effects. In the lower panel, we fix the parameter $R=2$ at the upper boundary $z=5H$. For instance, at the  frequency of $|\omega|=3.22h\Omega_K$, the growth rate decreases from $0.381h\Omega_K$ to $0.307h\Omega_K$ as the wavenumber $m$ increases from $0$ to $3$.} 

{Comparing the two panels of~Fig.~\ref{fig:non-axisymmetric}, the evident changes of unstable modes caused by BCs occur primarily for higher-order body modes, whereas non-axisymmetric effects demonstrate a global impact across the entire frequency range for both surface and body modes.
The non-axisymmetric effect is significant and can not be neglected.}
Though we do not yet have a satisfactory physical explanation for the $m$-dependence of VSI, \deleted{we plan to conduct further investigations on this issue.}
\added{our calculations indicate the term $-3mh\hat{k}_r$ in the variable $W=4imh^2-3mh\hat{k}_r$ plays an essential role for this dependence.} 

\section{Conclusion}

In this paper, we analyze the VSI with partially reflecting BCs for vertically global and radially local protoplanetary disks. The partially reflecting BCs are viewed as a linear combination of the {incoming and outgoing wave components}. By adjusting the relative magnitudes of the two wave components, we can investigate the response of the VSI to different BCs.

We find that the growth rate of VSI is closely related to the BCs, especially for higher-order breathing and corrugation modes whose frequency exceeds the threshold. The growth rate decreases when the boundary parameter, $R$, increases. In other words, the growth rate of the instability diminishes when the outgoing wave component dominates. 
As the outgoing component increases, the location of the maximum growth rate shifts toward a lower frequency. 

How waves propagate away from the surface of 
realistic protoplanetary disks is determined by the atmosphere of the disk \citep{1999ApJ...515..767O}. The fraction of the reflected and transmitted wave depends critically on the transition region between the disk and the atmosphere. In this paper, we use a free boundary parameter, $R$, to describe the wave behavior at this disk-atmosphere transition region. 
Whether the disk atmosphere enhances or suppresses the VSI depends on the characteristics of this transition region. 
Our current understanding of the physical conditions of protoplanetary disks and their atmosphere is still limited. It would be interesting to perform further studies on wave propagation in a more realistic disk-atmosphere transition region. 

The vertical velocity perturbation exhibits symmetric or anti-symmetric properties in the case of reflective BC at both the upper and lower boundaries \citep{nelson_linear_2013}. It is possible that the upper and lower disk-atmosphere transition regions have different physical properties. In this case, the waves would behave differently at the upper and lower boundaries, and the inherent symmetry (or anti-symmetry) would be broken. 

{We examine the growth rate of VSI with different vertical extents. 
As the vertical extent increases, the maximum growth rate of VSI increases as well. The finding is consistent with the discoveries made by \citet{2015MNRAS.450...21B}.} {Under partially reflecting BCs, this property remains unchanged.}
{In \citet{lin_cooling_2015}, the free BCs establish a zero Lagrangian pressure perturbation, consistent with a reflecting boundary with a phase difference. Moreover, the rigid BC ($\delta{u_z}=0$) they used aligns with our calculations with $R=-1$. We have also performed calculations with boundary parameters $R=1$ (reflecting BC) and $R=-1$. In both of these cases, the growth rates of VSI are found to be consistent.}

 {\deleted{The non-axisymmetric effect also has a significant impact on the VSI, which influences both the surface modes and the body modes}. \deleted{As the azimuthal wavenumber $m$ increases, the non-axisymmetric modes tend to suppress the VSI further.}  \added{We find that the non-axisymmetric modes are also unstable and they grow at a rate that decreases with the azimuthal wavenumber.} 
 In a non-axisymmetric scenario, the BCs favoring the outgoing wave component still have a suppression effect on the VSI.} 

MHD disk winds can potentially initiate outgoing flow in realistic PPDs. The presence of anti-symmetric magnetic field modes leads to one-sided characteristics in disk winds, which can further cause asymmetry in the outgoing flow of the disk. 
These factors will affect the development of VSI and may significantly influence planet formation and evolution. Realistic protoplanetary disks are magnetized. It would be interesting to take into account the effects of magnetic fields, incorporating our new treatment of BCs to explore the VSI in magnetized protoplanetary disks.

\section*{Acknowledgements}
\added{We are grateful for the anonymous referee's valuable comments and suggestions that improve the manuscript.}
This work has been supported by the National SKA Program of China (Grant No. 2022SKA 0120101) and the National Key R$\&$D Program of China (No. 2020YFC2201200), the science research grants from the China Manned Space Project (No. CMS-CSST-2021-B09, CMSCSST-2021-B12 and CMS-CSST-2021-A10), and opening fund of State Key Laboratory of Lunar and Planetary Sciences (Macau University of Science and Technology) (Macau FDCT Grant No. SKL-LPS(MUST)-2021-2023). C.Y. has been supported by the National Natural Science Foundation of China (Grant Nos. 11521303, 11733010, 11873103, and 12373071).

\section*{Data Availability}
The data underlying this article will be shared on reasonable request
to the corresponding author.



\bibliographystyle{mnras}
\bibliography{example} 



\appendix

\section{Derivation of the Boundary Conditions}
\label{appendix B}
In this appendix, we show the derivation of Equation~(\ref{eq:boundary}). According to the WKB approximation $y_1, y_2 \propto e^{i \hat{k}_z \hat{z}}$, we can readily know that ${d y_1}/{d \hat{z}} = i \hat{k}_z y_1$, ${d y_2}/{d \hat{z}} = i \hat{k}_z y_2 $. \added{Note that $\hat{k_z}$ here is a local constant associated with the boundaries}\footnote{\added{We adopt the Wentzel-Kramers-Brillouin (WKB) approximation, which is typically used for a semi-classical calculation in quantum mechanics in which the wave function is recast as an exponential function \citep{Leonard1968}. That is the theme of our treatment for the boundary condition. This method aims to find approximate solutions to linear ODEs with spatially varying coefficients. For the global vertical shear instability analysis, the relevant disk physical variables are supposed to be viewed as wave components with vertical wave number that varies with the vertical positions. But when it comes to the boundaries, we actually focus on the very regions that are spatially local. As a result, the variations of wave number are small and could be approximated as a local constant. In addition to quantum mechanics, such WKB approximations are also widely used in the astrophysical wave dynamics \citep{2000ApJ...533.1023L, Cui2024}}. }. Then Eq.~(\ref{eq: eigenfunc 1}) becomes
    \begin{equation}
        (A_{11} - i \hat{k}_z )y_1 + A_{12} y_2 = 0  \ , \
    A_{21} y_1 + (A_{22} - i \hat{k}_z ) y_2 = 0  \  .
    \label{eq: B1}
    \end{equation}
To ensure the existence of nontrivial solutions of $y_1$ and $y_2$, the matrix $\boldsymbol{A}$ must satisfy 

    \begin{equation}
	\begin{vmatrix}\boldsymbol{A} - i\hat{k}_z \boldsymbol{I}\end{vmatrix} = \begin{vmatrix} A_{11} - i \hat{k}_z & A_{12} \\
		A_{21} & A_{22} - i \hat{k}_z 
	\end{vmatrix} =0 \ .
    \label{eq: A2}
    \end{equation}
{The equation~(\ref{eq: A2}) can be rewritten as a quadratic equation for $i\hat{k}_z$ :}
    \begin{equation}
        \left(i\hat{k}_z\right)^2 -i\hat{k}_z\left(1+Q\right)\hat{z} -\Delta = 0 \ , \ 
    \label{eq: kz}
    \end{equation}
where $\Delta = -\hat{{\upsilon}}^2(\chi+U) +(\chi-1)\hat{z}^2\left[Q-U\right]$. 
This quadratic equation for $i \hat{k}_z$ can be readily solved. The two roots are 
\begin{equation}
    i \hat{k}_{z1} = \frac{1}{2}\left[ (1+Q)\hat{z} + i\sqrt{A_3-A_2\hat{z}^2}\right] \ , \
\end{equation}
\begin{equation}
    i \hat{k}_{z2} = \frac{1}{2}\left[ (1+Q)\hat{z} - i\sqrt{A_3-A_2\hat{z}^2}\right] \ . \
\end{equation}
    Note that there are two eigenvectors associated with the two eigenvalues of the matrix ${\boldsymbol{A}}$\footnote{As our convention, we might simply treat $\hat{k}_z$ as the eigenvalue of the matrix $\boldsymbol{A}$, though there is a difference between $\hat{k}_z$ and $\lambda$ induced by the imaginary unit.}. 
    According to the standard algorithm in linear algebra, the two eigenvectors associated with $i\hat{k}_{z1}$ and $i\hat{k}_{z2}$ can be written as：
    \begin{equation}
        {\boldsymbol{r}_1} = \begin{pmatrix}
            B_1\\-A_1\hat{z}+ i\sqrt{A_3-A_2\hat{z}^2}
        \end{pmatrix} 
        \ , \ 
        \boldsymbol{r}_2=\begin{pmatrix}
            B_1\\-A_1\hat{z}- i\sqrt{A_3-A_2\hat{z}^2}
        \end{pmatrix} \ ,
        \label{eq:B12}
    \end{equation}
     where
    \begin{equation}
        A_1=2\chi+Q-1 \ ,  \ 
        A_2 = \left(Q+1\right)^2+ 4\left(\chi-1\right)\left(Q-U\right) \ ,
    \end{equation}
    \begin{equation}
        A_3 = 4\hat{\upsilon}^2\left(\chi+U\right) \ , \  B_1=2{i\hat{\upsilon}}\left(\chi+U\right) \ ,
    \end{equation}
    respectively.  Note that these two eigenvectors represent waves with different group velocities. 
    
    The equation~(\ref{eq:23}) can be written explicitly as:
    \begin{equation}
        (a_1 +a_2)=y_1/B_1 \ ,    
    \end{equation}
    \begin{equation}
        (a_1 -a_2)=\frac{y_2+A_1 \hat{z}(a_1 +a_2)}{i\sqrt{A_3-A_2\hat{z}^2}}
        =\frac{y_2+A_1\hat{z}y_1 /B_1}{i\sqrt{A_3-A_2\hat{z}^2}} \ .
    \end{equation}
    With some simple manipulations, we can arrive at:
    \begin{equation}
        R \equiv \frac{a_2}{a_1}=\dfrac
	{\left[-{A_1}\hat{z}+i\sqrt{ {A_3}-{A_2}\hat{z}^2}\right] {y_1}-{B_1}{y_2}}	{\left[{A_1}\hat{z}+i\sqrt{{A_3}-{A_2}\hat{z}^2}\right] {y_1}+{B_1}{y_2}} \ ,
    \end{equation}
    which could be recast as a linear combination of $y_1$ and $y_2$, namely:
    \begin{equation}
        \left({A_1}\hat{z}+\frac{R-1}{R+1}i\sqrt{{A_3}-{A_2}\hat{z}^2}\right){y_1}+{B_1}{y_2}=0 \ .
    \end{equation}
This is exactly Equation~(\ref{eq:boundary}) in the main text. A larger $R$ indicates a greater outgoing component. Its non-axisymmetric counterpart, the Equation~(\ref{eq:boundary_m}) could be obtained in a similar manner.

Note that our BCs are quite general and the free BC (zero Lagrangian pressure perturbation, i.e., $\Delta P = 0$) in \citet{lin_cooling_2015} can be recovered with an appropriate value of $R_{\rm free}$. 
Numerical calculation shows that the modulus of $|R_{\rm free}| = 1$. The free BC actually means a reflecting boundary with a phase difference between the incoming and outgoing waves.

\section{Dispersion relation and Identification of Wave Propagation Direction}
\label{appendix A}
With the WKB approximation, the local dispersion relation is:
    \begin{equation}
c_4\hat{\upsilon}^4+c_3\hat{\upsilon}^3+c_2\hat{\upsilon}^2+c_1\hat{{\upsilon}}+c_0=0\ ,
    \label{eq: dispersion equation}
    \end{equation}
where coefficient{s} $c_0$, $c_1$, $c_2$, $c_3$ {and} $c_4$ are
    \begin{equation}
c_0=\hat{k}_z^2+i\left(Q+1\right)\hat{z}\hat{k}_z \ , \
        c_1=i\beta\left(\gamma-1\right)\left(Q-U\right)\hat{z}^2 \ ,
    \end{equation}
    \begin{equation}
        c_2=\beta^2\gamma^2c_0+i\beta\gamma{c_1}-U-1  \ , \ 
        c_3=-i\beta\left(\gamma-1\right) \ ,
    \end{equation}
    \begin{equation}
        c_4=-\beta^2\gamma\left[1+\gamma{U}\right] \ , 
    \end{equation}
    respectively. Note that this dispersion relation is equivalent to the Equation~(\ref{eq: kz}).  
    According to the definition of group velocity $v_g$=$\frac{\partial\upsilon}{\partial\hat{k}_z}$, 
    we know that the group velocity is：
    \begin{equation}
        v_g =-(\hat{\upsilon}^2\beta^2\gamma^2+1)
\left[2\hat{k}_z+i\left(Q+1\right)\hat{z}\right]/D \ .
        \label{eq: group velocity}
    \end{equation}
    where $D ={4c_4\hat{\upsilon}^3+3c_3\hat{\upsilon}^2+
        2c_2\hat{\upsilon}+c_1}$. 
    \added{The inverse function rule is applied during the calculation because $\hat{k}_z$ depends on $\upsilon$ and not the other way around.}
    
    Note that there are two roots of the quadratic equation (\ref{eq: kz}) for $\hat{k_z}$. Each root of $\hat{k_z}$ has a corresponding eigenvector. The specific form of each eigenvector is shown in Equation (\ref{eq:B12}). It is clear that, for the wave associated with $\boldsymbol{{r}}_2$, 
    the group velocity $v_g > 0$ (outgoing) at the outer boundary where $z=5H$. 
    For the wave associated with  $\boldsymbol{{r}}_1$, 
    the group velocity $v_g<0$ (incoming) at outer boundary $z=5H$. \added{The wave assignment is general for all eigenfunctions and independent of the chosen parameters. It also holds in general for the case of $m\neq 0$.}


\bsp	
\label{lastpage}

\end{CJK*}
\end{document}